\pdfoutput=1
\documentclass[preprint,preprintnumbers,amsmath,amssymb,floatfix,endfloats*]{revtex4}

\usepackage{graphicx}
\usepackage{dcolumn}
\usepackage{bm}
\usepackage{amsfonts}

\DeclareMathAlphabet{\mathsfsl}{OT1}{cmr}{bx}{it}
\begin{document}
\title{Alternating shear orientation during cyclic loading facilitates yielding in amorphous materials}
\author{Nikolai V. Priezjev$^{1,2}$}
\affiliation{$^{1}$Department of Mechanical and Materials
Engineering, Wright State University, Dayton, OH 45435}
\affiliation{$^{2}$National Research University Higher School of
Economics, Moscow 101000, Russia}
\date{\today}
\begin{abstract}

The influence of alternating shear orientation and strain amplitude
of cyclic loading on yielding in amorphous solids is investigated
using molecular dynamics simulations. The model glass is represented
via a binary mixture that was rapidly cooled well below the glass
transition temperature and then subjected to oscillatory shear
deformation. It was shown that periodic loading at strain amplitudes
above the critical value first induces structural relaxation via
irreversible displacements of clusters of atoms during a number of
transient cycles, followed by an increase in potential energy due to
the formation of a system-spanning shear band. Upon approaching the
critical strain amplitude from above, the number of transient cycles
required to reach the yielding transition increases. Interestingly,
it was found that when the shear orientation is periodically
alternated in two or three dimensions, the number of transient
cycles is reduced but the critical strain amplitude remains the same
as in the case of periodic shear along a single plane. After the
yielding transition, the material outside the shear band continues
strain-induced relaxation, except when the shear orientation is
alternated in three dimensions and the glass is deformed along the
shear band with the imposed strain amplitude every third cycle.

\vskip 0.5in


\end{abstract}

\maketitle

\section{Introduction}

Understanding the relationship between the local atomic structure of
amorphous alloys and their mechanical and physical properties is
important for various structural, biomedical, and environmental
applications~\cite{Qiao19,ZhengBio16,catalytic19}.  It is well
accepted that in contrast to crystalline materials, where plastic
deformation is governed by motion of topological line defects, or
dislocations, the elementary plastic event in amorphous materials
involves a collective rearrangement of small group of atoms, or a
shear transformation~\cite{Spaepen77,Argon79}. The lack of
crystalline order in metallic glasses results in relatively high
yield strength, and, at the same time, if brought to a relaxed
state, glasses can fail via sudden formation of shear bands where
strain becomes localized along narrow layers~\cite{Ma11}.   On the
other hand, metallic glasses can be made more ductile if they are
mechanically or thermally rejuvenated or, alternatively, formed by
rapid cooling from the liquid state. In the past, a number of
thermomechanical processing methods, such as cold rolling,
high-pressure torsion, thermal cycling, elastostatic loading, and
irradiation, were developed to rejuvenate glasses and improve
plasticity~\cite{Greer16}.  Despite recent advances, however, the
structural relaxation and critical behavior in glasses during
time-periodic mechanical deformation are yet to be fully understood.

\vskip 0.05in

In the last few years, the processes of structural relaxation,
rejuvenation and yielding in amorphous materials under cyclic
loading were extensively studied using atomistic
simulations~\cite{Priezjev13,Sastry13,Reichhardt13, Priezjev14,
Priezjev15, IdoNature15, Priezjev16, Kawasaki16, Priezjev16a,
Sastry17, Priezjev17, OHern17, Priezjev18, Priezjev18a,
NVP18strload, PriMakrho05, PriMakrho09, Sastry19band, Peng19,
PriezSHALT20, Pelletier20, Ido2020, Priez19ba, Jana19, KawBer19,
BhaSastry19}.  In general, the results depend sensitively on the
simulation model, preparation history, details of deformation
protocol, and system size. In the athermal limit, using quasistatic
oscillatory shear deformation protocol, it was shown that following
a number of periods, a disordered system reaches the so-called
`limit cycle' where the trajectory of each atom becomes exactly
reversible after one or more cycles, despite that during each
subyield cycle large clusters of atoms can undergo cooperative
displacements~\cite{Reichhardt13,IdoNature15}. Interestingly, it was
recently shown that highly stable glasses, which were produced
either using the swap Monte Carlo algorithm~\cite{KawBer19} or via
mechanical annealing~\cite{BhaSastry19}, can be reversibly deformed
over a relatively broad range of strain amplitudes.  When the strain
amplitude is above a critical value, the yielding transition usually
occurs after a number of transient cycles and it is accompanied by
the formation of a shear band across the
system~\cite{Sastry17,Priezjev17,Priezjev18a,PriMakrho09,Sastry19band,Priez19ba}.
While during startup continuous shear deformation of well annealed
glasses, the position of strain localization cannot be determined
until the yielding strain~\cite{Priez19star}, the location of a
shear band during time periodic deformation can be identified at
least several cycles before the yielding
transition~\cite{Priez19ba}. Furthermore, it was previously
demonstrated that mechanical annealing during small-amplitude cyclic
shear can be accelerated by periodically alternating shear
orientation in two or three spatial dimensions~\cite{PriezSHALT20}.
It remains unclear, however, how the yielding transition and shear
band formation are affected by cyclic deformation with alternating
orientation of the shear plane near the critical strain amplitude.

\vskip 0.05in

In this paper, the effects of alternating shear orientation and
strain amplitude on yielding and relaxation in binary glasses during
cyclic loading are studied using molecular dynamics simulations.
The binary mixture is initially rapidly cooled deep into the glass
phase and then subjected to periodic shear applied either along a
single plane or alternated in two or three dimensions. It will be
shown that the number of transient cycles before yielding is reduced
when an additional shear orientation is introduced in the
deformation protocol at a given strain amplitude.  Moreover, the
critical strain amplitude, below which the glass undergoes
relaxation towards lower potential energy states, remains the same
for periodic shear along a single or alternating planes.

\vskip 0.05in

The rest of the paper is organized as follows. The details of the
molecular dynamics simulation model and oscillatory shear
deformation protocols are described in the next section. The time
dependence of the potential energy and shear stress as well as the
analysis of nonaffine displacements during cyclic loading are
presented in section\,\ref{sec:Results}.  The results are briefly
summarized in the last section.

\section{Molecular dynamics simulations}
\label{sec:MD_Model}

The disordered solid is modeled via the binary Lennard-Jones (LJ)
mixture where the interaction between different types of atoms is
strongly non-additive, thus preventing crystallization below the
glass transition temperature~\cite{KobAnd95}. This model was first
developed and its properties were thoroughly examined by Kob and
Andersen (KA) using molecular dynamics simulations~\cite{KobAnd95}.
The parametrization of the KA model is similar to the description of
the amorphous metal-metalloid alloy $\text{Ni}_{80}\text{P}_{20}$
originally studied by Weber and Stillinger~\cite{Weber85}. In the KA
model, any two atoms interact via the LJ potential given by
\begin{equation}
V_{\alpha\beta}(r)=4\,\varepsilon_{\alpha\beta}\,\Big[\Big(\frac{\sigma_{\alpha\beta}}{r}\Big)^{12}\!-
\Big(\frac{\sigma_{\alpha\beta}}{r}\Big)^{6}\,\Big],
\label{Eq:LJ_KA}
\end{equation}
with the following parameters: $\varepsilon_{AA}=1.0$,
$\varepsilon_{AB}=1.5$, $\varepsilon_{BB}=0.5$, $\sigma_{AA}=1.0$,
$\sigma_{AB}=0.8$, $\sigma_{BB}=0.88$, and
$m_{A}=m_{B}$~\cite{KobAnd95}. The system consists of $48\,000$ of
atoms of type $A$ and $12\,000$ $B$ type atoms, and the total number
of atoms is fixed to $60\,000$ throughout the study.  To speed up
computations, the cutoff radius is set to
$r_{c,\,\alpha\beta}=2.5\,\sigma_{\alpha\beta}$.  The numerical
results are reported using the reduced LJ units of length, mass, and
energy, as follows: $\sigma=\sigma_{AA}$, $m=m_{A}$, and
$\varepsilon=\varepsilon_{AA}$.  The equations of motion were
integrated in parallel using the velocity Verlet algorithm with the
time step $\triangle t_{MD}=0.005\,\tau$, where
$\tau=\sigma\sqrt{m/\varepsilon}$ is the LJ
time~\cite{Allen87,Lammps}.

\vskip 0.05in


We next briefly describe the preparation procedure and the
deformation protocol. After equilibration at the temperature
$T_{LJ}=1.0\,\varepsilon/k_B$ and density
$\rho=\rho_A+\rho_B=1.2\,\sigma^{-3}$, the binary mixture was
rapidly cooled with the rate $10^{-2}\varepsilon/k_{B}\tau$ to the
low temperature $T_{LJ}=0.01\,\varepsilon/k_B$ at constant volume.
For reference, the critical temperature of the KA model at
$\rho=1.2\,\sigma^{-3}$ is $T_c=0.435\,\varepsilon/k_B$, where $k_B$
is the Boltzmann constant~\cite{KobAnd95}. In our setup, periodic
boundary conditions were applied, and the temperature was regulated
via the Nos\'{e}-Hoover thermostat~\cite{Allen87,Lammps}. The linear
size of the periodic box is $L=36.84\,\sigma$.

\vskip 0.05in


After rapid cooling to the temperature
$T_{LJ}=0.01\,\varepsilon/k_B$, the binary glass was periodically
sheared with the period $T=5000\,\tau$ at constant volume, as
follows:
\begin{equation}
\gamma(t)=\gamma_0\,\text{sin}(2\pi t/T),
\label{Eq:shear}
\end{equation}
where $\gamma_0$ is the strain amplitude in the vicinity of the
yielding point, $0.065\leqslant\gamma_0\leqslant0.070$, and the
oscillation frequency is
$\omega=2\pi/T=1.26\times10^{-3}\,\tau^{-1}$. The deformation was
applied either along a single plane (the $xz$ plane), or alternated
between two planes (the $xz$ and $yz$ planes), or alternated in all
three directions (\textit{i.e.}, along the $xz$, $yz$, and $xy$
planes). The deformation protocols are the same as in the recent
study on mechanical annealing under periodic
shear~\cite{PriezSHALT20}. In the present study, the data were
acquired only for one realization of disorder due to the
considerable computational demands. For example, a typical
production run at a given strain amplitude during 2600 cycles
required about 70 days using 40 processors.

\section{Results}
\label{sec:Results}


It is well realized that periodic deformation of rapidly quenched
glasses can lead to progressively lower energy states if the
temperature is sufficiently below the glass transition temperature
and the strain amplitude is smaller than a critical
value~\cite{Lacks04,Sastry13,Sastry17,Priezjev18,Priezjev18a}. The
structural relaxation originates from irreversible rearrangements of
groups of atoms, which can relocate to nearby minima in the
potential energy landscape during strain~\cite{Lacks04}. In the
recent study, it was demonstrated that evolution to more relaxed
states is accelerated when shear orientation is alternated in two or
three spatial dimensions~\cite{PriezSHALT20}.   During alternating
loading, the potential energy landscape is periodically deformed in
different directions and groups of atoms can rearrange to deeper
energy minima.   In the following analysis, the influence of the
deformation protocol with alternating shear orientation on yielding
and shear band formation is examined near the critical strain
amplitude.

\vskip 0.05in

\begin{figure}[t]
\includegraphics[width=12.0cm,angle=0]{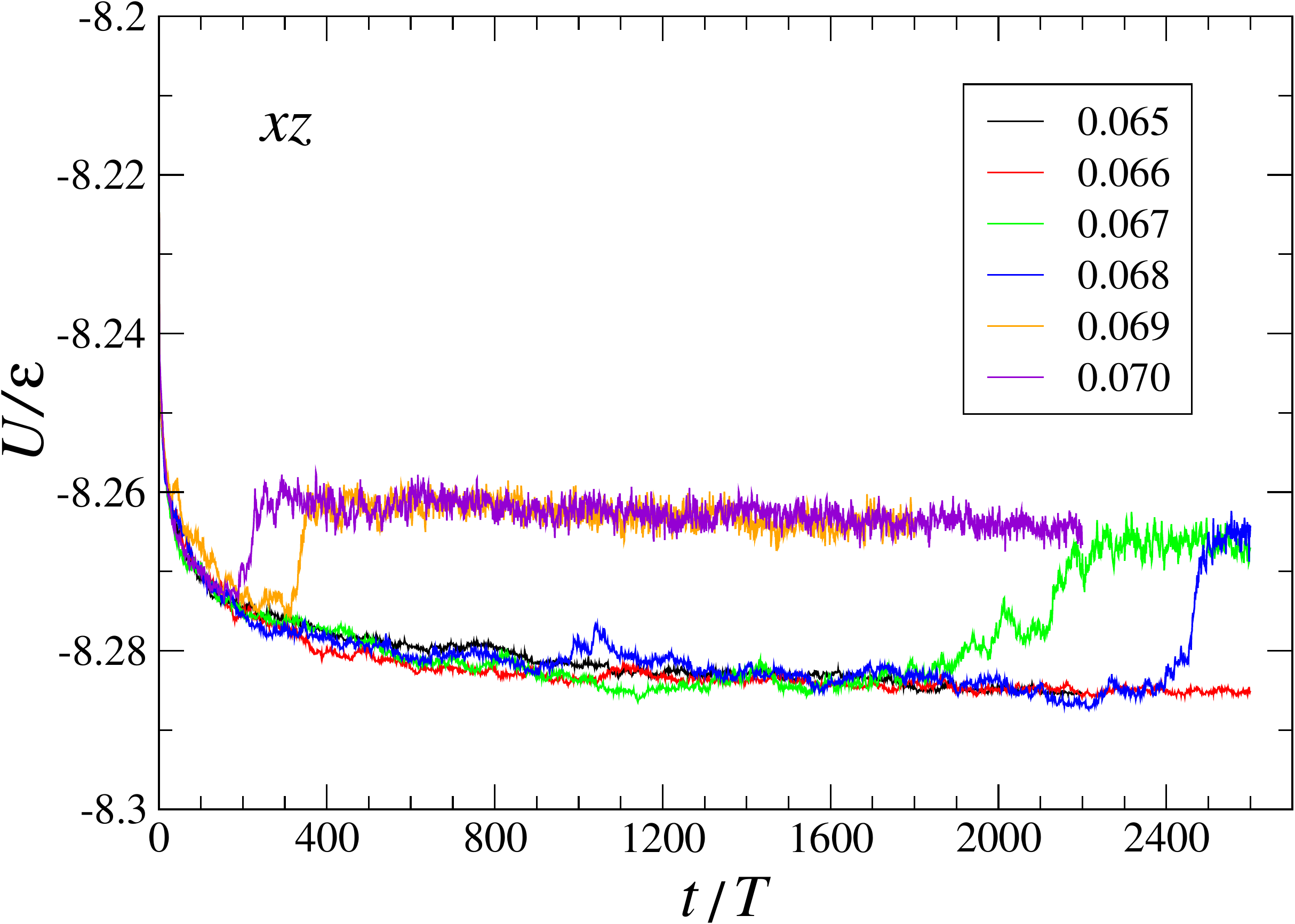}
\caption{(Color online) The potential energy minima as a function of
the cycle number during periodic shear along the $xz$ plane. The
strain amplitudes are indicated in the inset. The period of
oscillation is $T=5000\,\tau$ and the temperature is
$T_{LJ}=0.01\,\varepsilon/k_B$.}
\label{fig:poten_xz_065_070}
\end{figure}

The time dependence of the potential energy minima during periodic
shear along a single plane (the $xz$ plane) is presented in
Fig.\,\ref{fig:poten_xz_065_070} for the strain amplitudes
$0.065\leqslant\gamma_0\leqslant0.070$. Here, the data are reported
at the end of each cycle, when the net strain is zero, and the time
is expressed in terms of the oscillation period, $T=5000\,\tau$. It
can be observed that in all cases, the potential energy initially
rapidly decreases from $U\approx-8.20\,\varepsilon$, which
corresponds to the state right after cooling with the rate
$10^{-2}\varepsilon/k_{B}\tau$ to $T_{LJ}=0.01\,\varepsilon/k_B$,
down to $U\approx-8.274\,\varepsilon$ at $t=200\,T$. The rapid decay
of the potential energy is expected, since the glass is initially
quenched at a relatively high cooling rate, and, therefore, its
local structure contains many high-energy clusters of atoms that are
prone to rearrangement under external perturbation. Upon further
cycling, the potential energy continues to decrease at a slower rate
when the strain amplitude is below the critical value,
\textit{i.e.}, $\gamma_0\leqslant0.066$. These results are
consistent with previous findings on mechanical annealing of
periodically driven glasses at
zero~\cite{Reichhardt13,Sastry13,Sastry17} and
finite~\cite{Priezjev18,Priezjev18a,PriezSHALT20,Jana19}
temperatures.

\vskip 0.05in


By contrast, periodic shear deformation at higher strain amplitudes,
$0.067\leqslant\gamma_0\leqslant0.070$, leads to the yielding
transition after a number of cycles, as shown in
Fig.\,\ref{fig:poten_xz_065_070}. It can be seen that at relatively
large strain amplitudes, $\gamma_0=0.069$ and $0.070$, the abrupt
increase in the potential energy occurs after about 300 and 200
cycles, respectively. Upon approaching the critical value of the
strain amplitude, $\gamma_0=0.067$, the number of transient cycles
increases significantly.  Somewhat unexpectedly, the yielding
transition at the strain amplitude $\gamma_0=0.068$ is delayed by
about 400 cycles with respect to the case $\gamma_0=0.067$, although
the transient response is subject to large fluctuations
(\textit{e.g.}, notice a local spike at $t\approx1000\,T$ in the
blue curve in Fig.\,\ref{fig:poten_xz_065_070}).   It can be further
observed that after the potential energy abruptly increases due to
the formation of a shear band (to be discussed below), the region
outside the shear band continues annealing, which is reflected in
slightly negative slope of the potential energy as a function of the
cycle number. This effect appears because strain is localized within
the shear band, and the material outside the shear band is deformed
at a strain amplitude smaller than the critical value. Similar
results for the strain localization and relaxation of the solid
phase as a function of accumulated strain were reported during
cyclic athermal quasistatic deformation~\cite{Sastry19band}.
Interestingly, the potential energy curves in
Fig.\,\ref{fig:poten_xz_065_070} essentially follow either one of
the two energy levels (except during the yielding transition) for
the strain amplitudes near the critical value
$0.065\leqslant\gamma_0\leqslant0.070$.

\vskip 0.05in

\begin{figure}[t]
\includegraphics[width=12.0cm,angle=0]{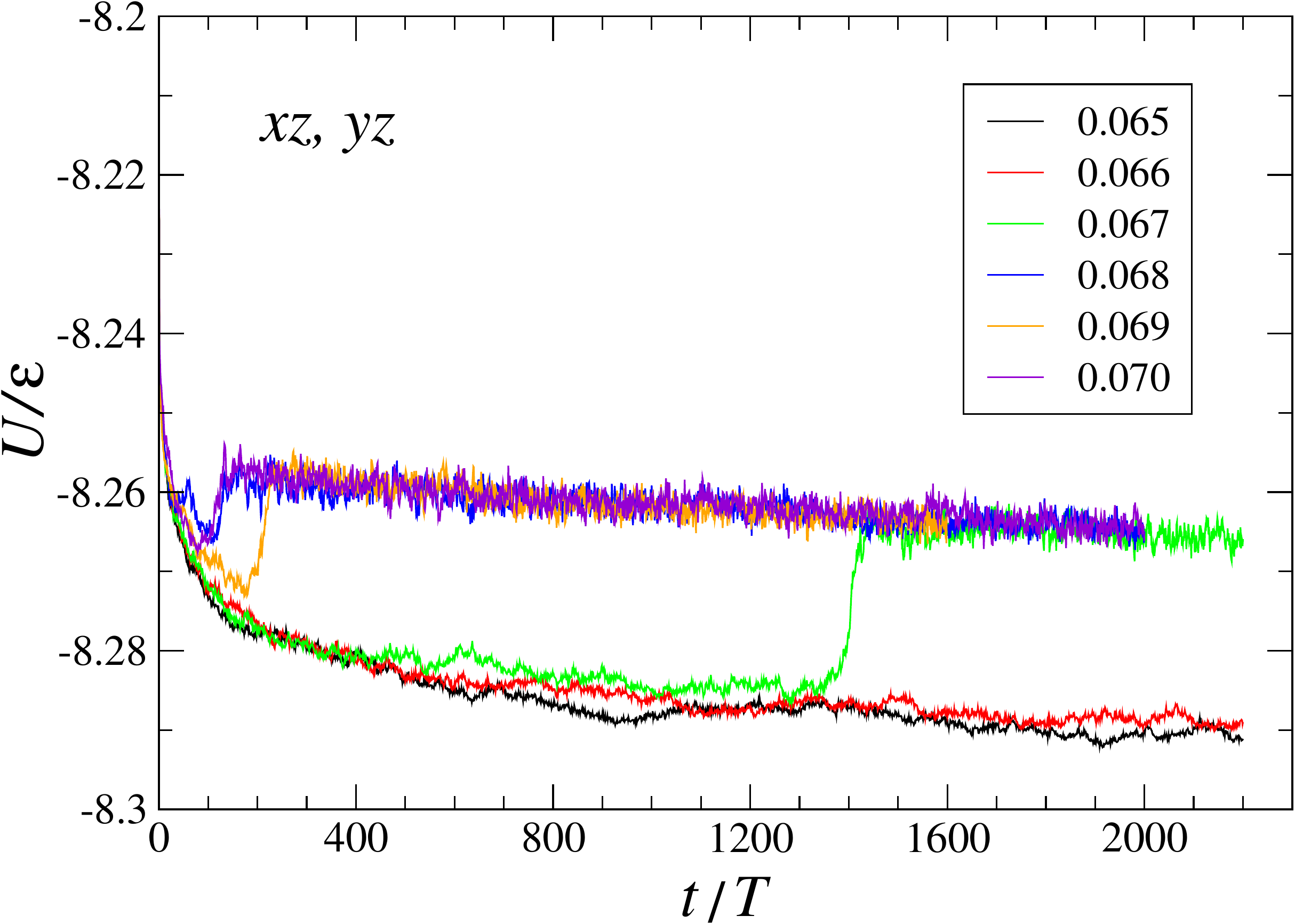}
\caption{(Color online) The variation of the potential energy minima
versus the cycle number during alternating shear along the $xz$ and
$yz$ planes. The strain amplitudes are $\gamma_0=0.065$ (black),
$0.066$ (red), $0.067$ (green), $0.068$ (blue), $0.069$ (orange),
and $0.070$ (velvet). The oscillation period is $T=5000\,\tau$.}
\label{fig:poten_xz_yz_065_070}
\end{figure}

The results for the potential energy versus cycle number for the
deformation protocol where the orientation of the shear plane is
alternated between the $xz$ and $yz$ planes are shown in
Fig.\,\ref{fig:poten_xz_yz_065_070}. The behavior is similar to the
case of periodic shear along a single plane, except that the sharp
yielding transition at the critical strain amplitude
$\gamma_0=0.067$ occurs after about $1400\,T$ rather than a gradual
crossover after $2000\,T$ at $\gamma_0=0.067$ reported in
Fig.\,\ref{fig:poten_xz_065_070}.  Note also that the number of
transient cycles decreases drastically for the strain amplitude
$\gamma_0=0.068$, \textit{i.e.}, about 100 cycles for alternating
loading in Fig.\,\ref{fig:poten_xz_yz_065_070} and about 2500 cycles
for periodic shear along a single plane in
Fig.\,\ref{fig:poten_xz_065_070}.

\vskip 0.05in

\begin{figure}[t]
\includegraphics[width=12.0cm,angle=0]{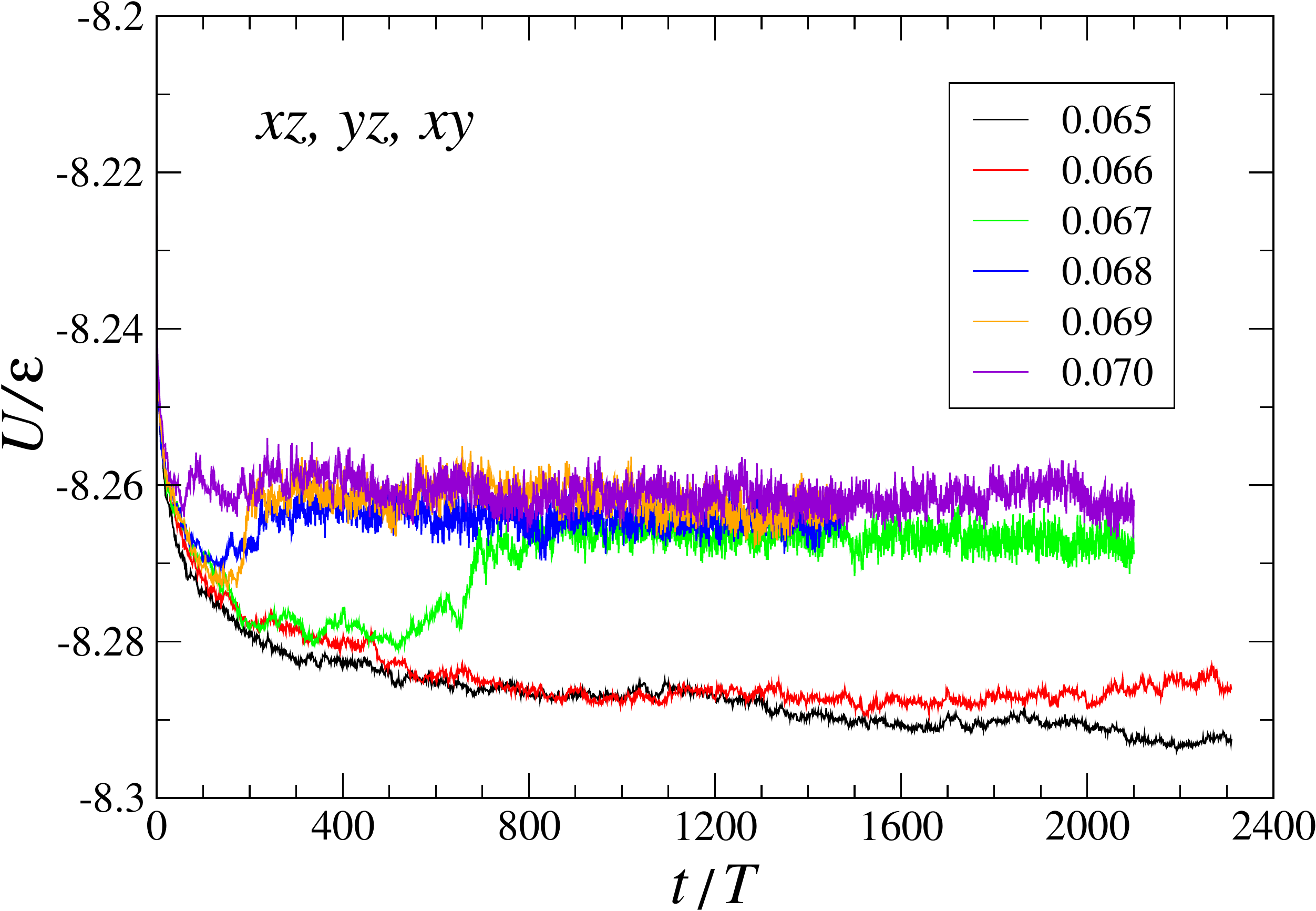}
\caption{(Color online) The potential energy at the end of each
cycle during alternating shear along the $xz$, $yz$, and $xy$
planes. The values of the strain amplitude are listed in the legend.
The time is measured in oscillation periods, $T=5000\,\tau$. }
\label{fig:poten_xz_yz_xy_065_070}
\end{figure}

Furthermore, the data for the alternating shear along the $xz$,
$yz$, and $xy$ planes are presented in
Fig.\,\ref{fig:poten_xz_yz_xy_065_070} for the same strain
amplitudes, $0.065\leqslant\gamma_0\leqslant0.070$. There are two
main differences from the previous cases. First, the yielding
transition at the critical strain amplitude $\gamma_0=0.067$ occurs
much sooner than for the other deformation protocols, \textit{i.e.},
already after about 600 cycles. Second, after a number of transient
cycles, a single shear band is formed along one of the planes for
loading at the strain amplitudes
$0.067\leqslant\gamma_0\leqslant0.070$. It means that one of the
three shear orientations coincides with the plane parallel to the
shear band, and, therefore, the material outside the shear band is
strained at the imposed $\gamma_0$ during one of the cycles, while
the effective strain amplitude is reduced during the other two
cycles. As a result, the annealing outside the shear band is
suppressed (notice nearly constant energy levels after the
transition), and the plateau levels for $t\gtrsim800\,T$ in
Fig.\,\ref{fig:poten_xz_yz_xy_065_070} become progressively higher
with increasing strain amplitude in the range
$0.067\leqslant\gamma_0\leqslant0.070$.

\vskip 0.05in

\begin{figure}[t]
\includegraphics[width=12.0cm,angle=0]{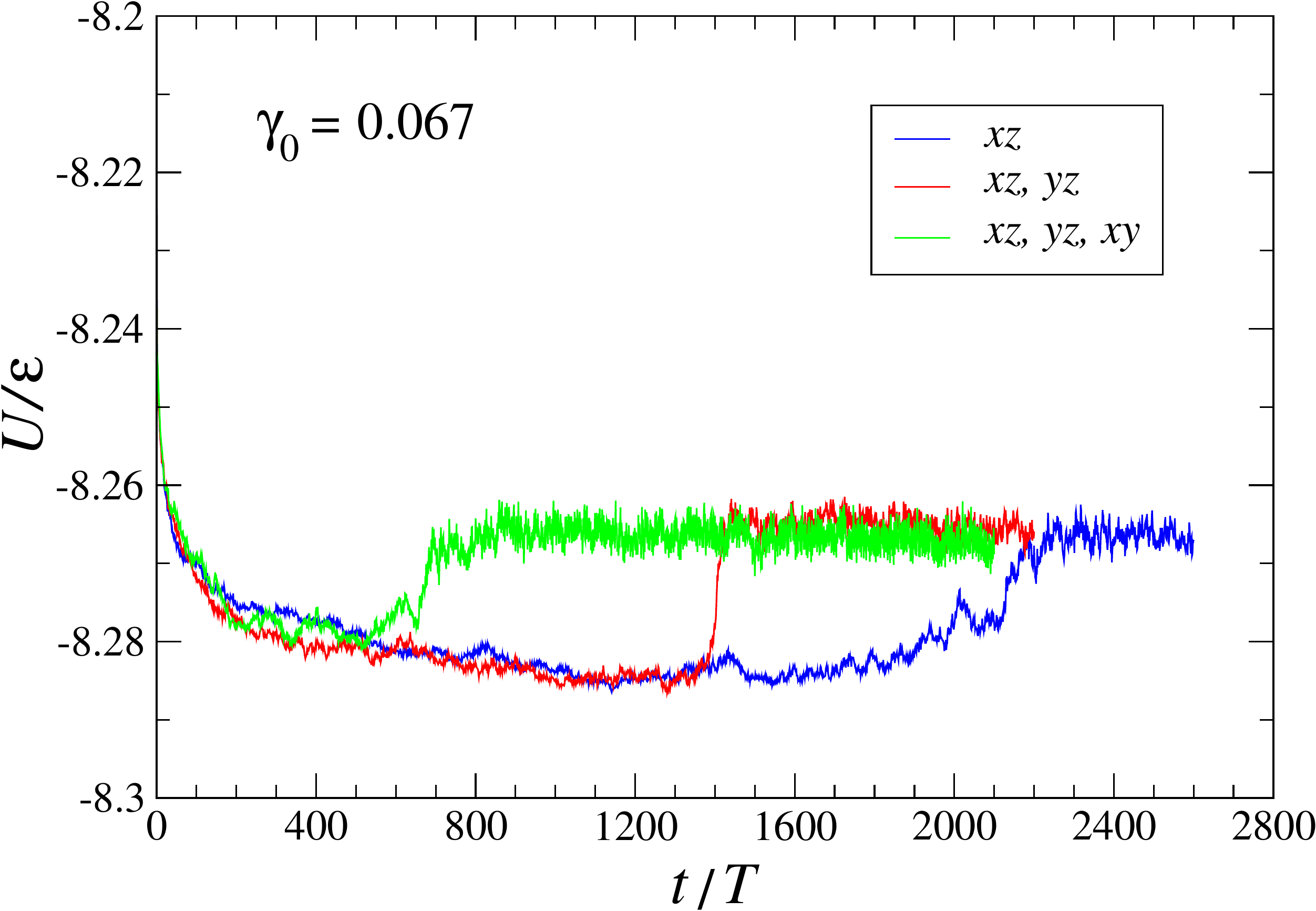}
\caption{(Color online) The potential energy minima as a function of
time for the strain amplitude $\gamma_0=0.067$. The deformation
protocols are: (1) periodic shear along the $xz$ plane, (2)
alternating shear along the $xz$ and $yz$ planes, and (3)
alternating shear along the $xz$, $yz$, and $xy$ planes. The same
data for $\gamma_0=0.067$ as in
Figs.\,\ref{fig:poten_xz_065_070}--\ref{fig:poten_xz_yz_xy_065_070}.}
\label{fig:poten_amp067_3prs}
\end{figure}

In order to facilitate comparison of the results for different
deformation protocols, the potential energy curves are replotted in
Fig.\,\ref{fig:poten_amp067_3prs} for the critical strain amplitude
$\gamma_0=0.067$.  As is evident, the potential energies are roughly
the same up to the yielding transition, and the number of shear
cycles to reach the transition point is reduced for protocols where
the shear orientation is periodically alternated in two and three
dimensions. In the context of the previous study on mechanical
annealing of periodically deformed glasses, where it was shown that
relaxation is accelerated with each additional alternation of the
shear orientation~\cite{PriezSHALT20}, it is important to note that
the critical value of the strain amplitude, $\gamma_0=0.067$,
remains unchanged for different deformation protocols. We comment,
however, that this conclusion is based on a limited number of cycles
at the strain amplitude $\gamma_0=0.066$, as reported in
Figs.\,\ref{fig:poten_xz_065_070}-\ref{fig:poten_xz_yz_xy_065_070}.


\vskip 0.05in

\begin{figure}[t]
\includegraphics[width=12.0cm,angle=0]{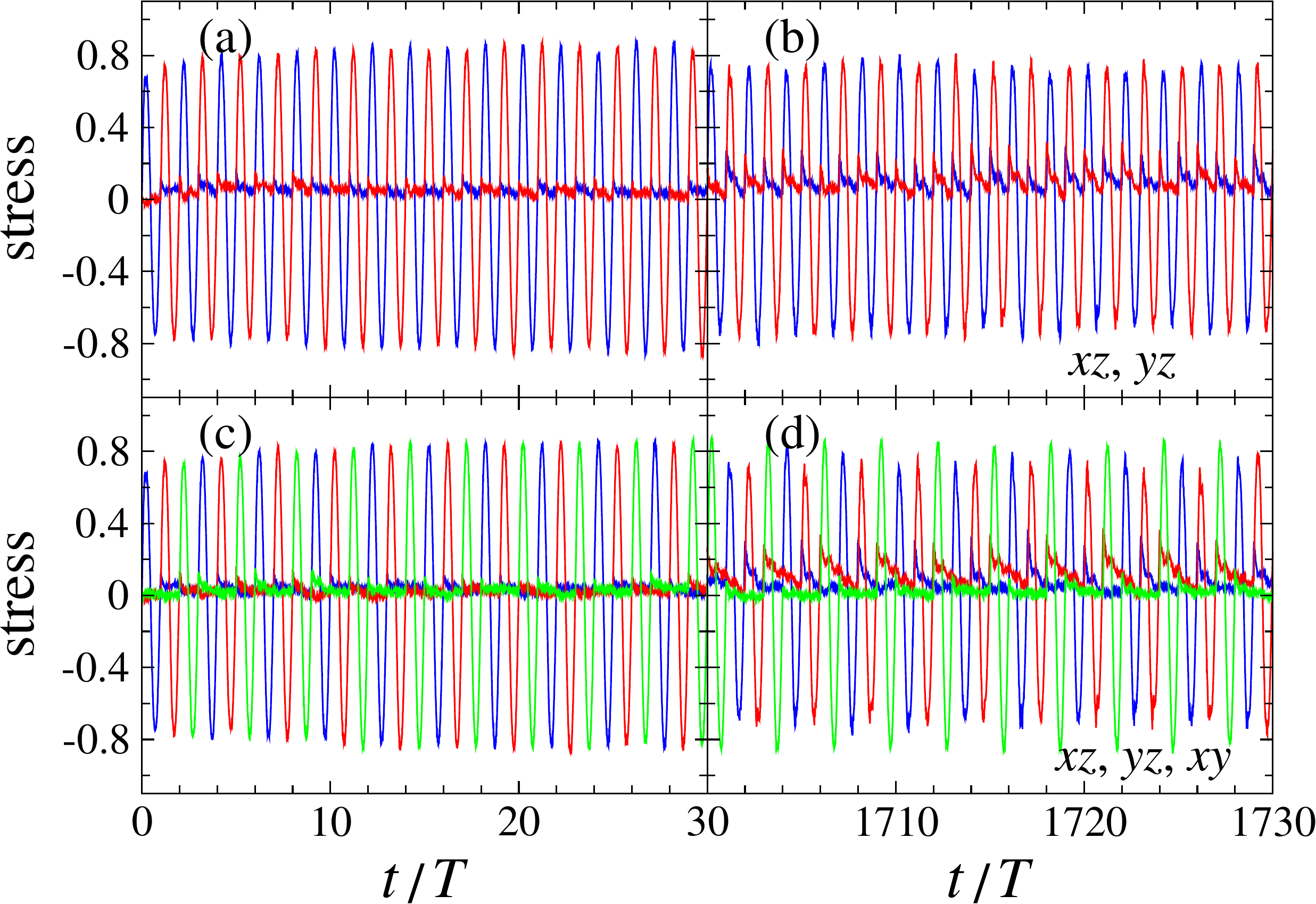}
\caption{(Color online) The time dependence of the shear stress (in
units of $\varepsilon\sigma^{-3}$) for the strain amplitude
$\gamma_0=0.067$. The upper panels show the data for alternating
shear along the $xz$ and $yz$ planes, and the lower panels are for
the alternating shear along the $xz$, $yz$, and $xy$ planes. The
data for the planes of shear are denoted by the blue ($xz$), red
($yz$), and green ($xy$) colors. The period is $T=5000\,\tau$.  }
\label{fig:stress_amp067_2prs}
\end{figure}

The variation of shear stress along the $xz$, $yz$, and $xy$ planes
during alternating loading is shown in
Fig.\,\ref{fig:stress_amp067_2prs} for the strain amplitude
$\gamma_0=0.067$. The data are presented during the first 30 cycles
after thermal annealing and during 30 cycles after the yielding
transition (see Fig.\,\ref{fig:poten_amp067_3prs}). It can be
observed in Fig.\,\ref{fig:stress_amp067_2prs}\,(a,\,c) that for
both deformation protocols, the stress amplitude initially
increases, as the glass becomes more relaxed, and it saturates to a
nearly constant level. These results are similar to the stress
variation during periodic shear along a single plane of rapidly
annealed binary glasses~\cite{Priezjev18,Priezjev18a}. By contrast,
the plastic flow within a shear band causes stress-strain
hysteresis, and, consequently, the shear stress at the end of each
cycle remains finite, and it approaches zero during the following
one or two cycles, as shown in
Fig.\,\ref{fig:stress_amp067_2prs}\,(b,\,d). The exception from this
behavior is the deformation \textit{along} the shear band in the
case of alternating shear in three dimensions; see the $xy$ stress
component (the green curve) in
Fig.\,\ref{fig:stress_amp067_2prs}\,(d).  Note also that the stress
amplitude during deformation along the $xy$ plane is significantly
larger than the amplitudes along the other directions, which are
determined by the maximum stress within the shear band.

\vskip 0.05in


The collective rearrangements of atoms during the relaxation stage
as well as the formation of a shear band can be visualized by
computing the so-called nonaffine displacements. We recall that the
nonaffine displacement of an atom is defined with respect to its
neighbors by using the matrix $\mathbf{J}_i$, which transforms the
positions of neighboring atoms during the time interval $\Delta t$
and at the same time minimizes the following quantity~\cite{Falk98}:
\begin{equation}
D^2(t, \Delta t)=\frac{1}{N_i}\sum_{j=1}^{N_i}\Big\{
\mathbf{r}_{j}(t+\Delta t)-\mathbf{r}_{i}(t+\Delta t)-\mathbf{J}_i
\big[ \mathbf{r}_{j}(t) - \mathbf{r}_{i}(t)    \big] \Big\}^2,
\label{Eq:D2min}
\end{equation}
where the summation is performed over atoms within a sphere of
radius $1.5\,\sigma$ and centered at $\mathbf{r}_{i}(t)$. It was
previously found that the nonaffine measure is particularly well
suited for identification of localized shear transformations in
quiescent and deformed disordered
solids~\cite{Falk98,Behringer08,Wang11}. More recently, the analysis
of nonaffine displacements was used to elucidate the structural
relaxation dynamics and yielding during time periodic
deformation~\cite{Priezjev16,Priezjev16a,Priezjev17,Priezjev18,
Priezjev18a,NVP18strload,PriezSHALT20,Priez19ba,Jana19} and thermal
processing~\cite{Priez19one,Priez19tcyc,Priez19T2000,Priez19T5000,PriezELAST19}
of binary glasses.

\vskip 0.05in

\begin{figure}[t]
\includegraphics[width=12.0cm,angle=0]{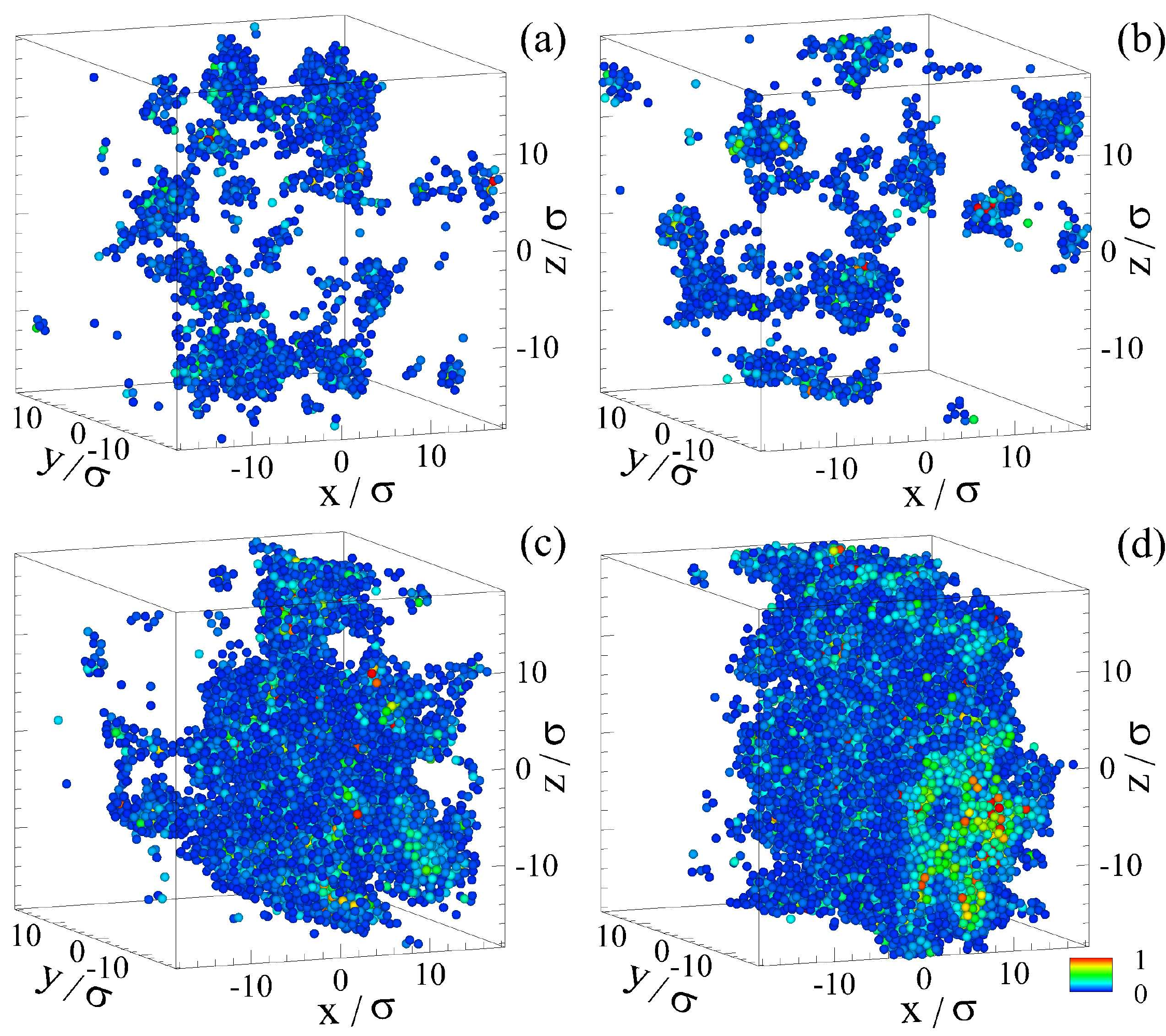}
\caption{(Color online) A sequence of snapshots during periodic
shear along the $xz$ plane with the strain amplitude
$\gamma_0=0.067$. The nonaffine measure is (a) $D^2(400\,T,
T)>0.04\,\sigma^2$, (b) $D^2(1200\,T, T)>0.04\,\sigma^2$, (c)
$D^2(2000\,T, T)>0.04\,\sigma^2$, and (d) $D^2(2400\,T,
T)>0.04\,\sigma^2$. The colorcode for $D^2$ is specified in the
legend. The oscillation period is $T=5000\,\tau$. The atoms are not
shown to scale. }
\label{fig:snapshots_xz_amp067}
\end{figure}
\begin{figure}[t]
\includegraphics[width=12.0cm,angle=0]{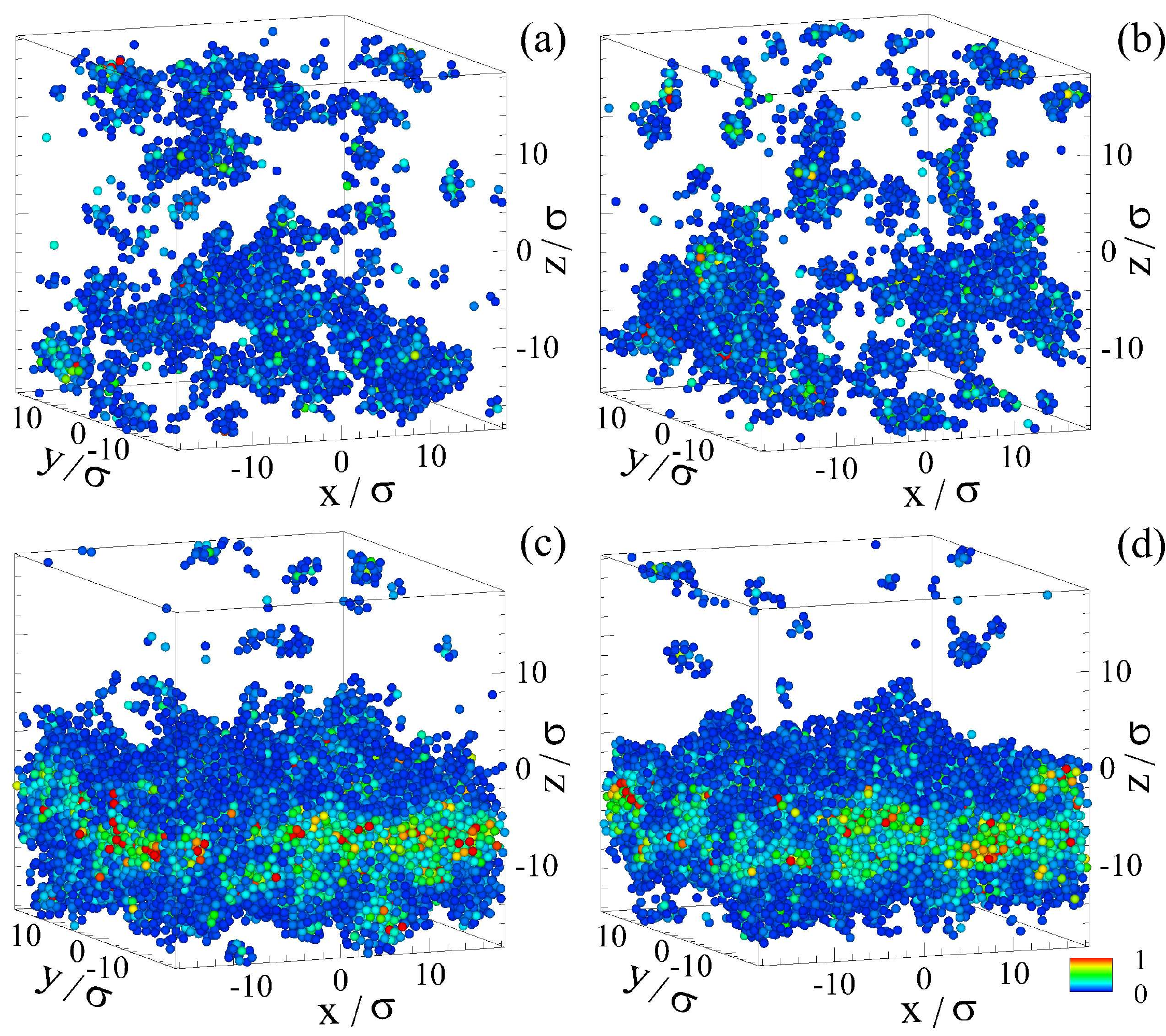}
\caption{(Color online) The snapshots of atomic configurations
during alternating shear along the $xz$ and $yz$ planes with the
strain amplitude $\gamma_0=0.067$. The nonaffine measure is (a)
$D^2(400\,T, T)>0.04\,\sigma^2$, (b) $D^2(1200\,T,
T)>0.04\,\sigma^2$, (c) $D^2(1600\,T, T)>0.04\,\sigma^2$, and (d)
$D^2(2000\,T, T)>0.04\,\sigma^2$. The colorcode for $D^2$ is defined
in the legend. }
\label{fig:snapshots_xz_yz_amp067}
\end{figure}
\begin{figure}[t]
\includegraphics[width=12.0cm,angle=0]{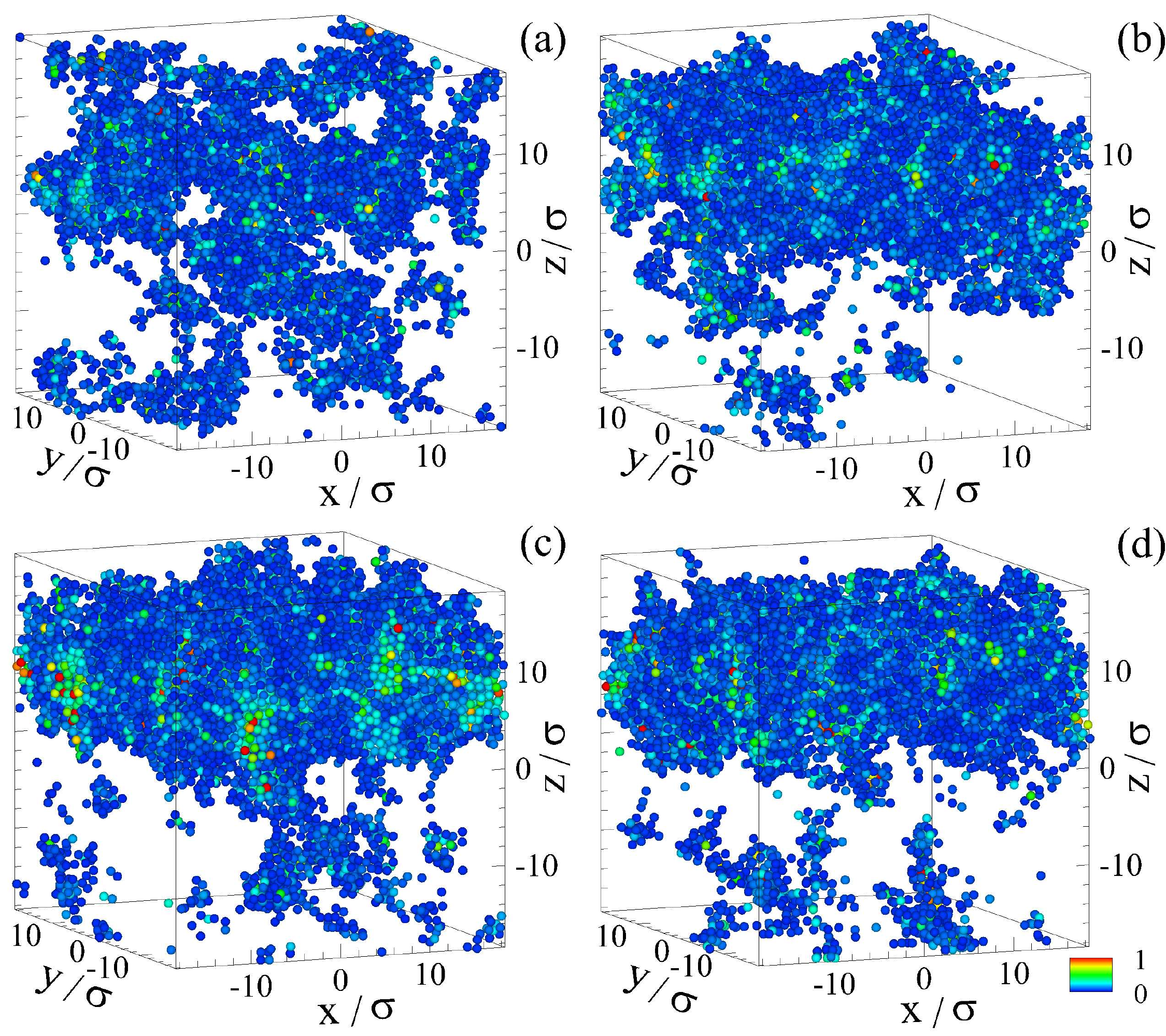}
\caption{(Color online) The snapshots of the system during
alternating shear along the $xz$, $yz$, and $xy$ planes with the
strain amplitude $\gamma_0=0.067$. The nonaffine measure is (a)
$D^2(400\,T, T)>0.04\,\sigma^2$, (b) $D^2(800\,T,
T)>0.04\,\sigma^2$, (c) $D^2(1200\,T, T)>0.04\,\sigma^2$, and (d)
$D^2(2000\,T, T)>0.04\,\sigma^2$. The colorcode for $D^2$ is shown
in the legend. }
\label{fig:snapshots_xz_yz_xy_amp067}
\end{figure}

The sequences of snapshots of atomic configurations for different
deformation protocols at the critical strain amplitude
$\gamma_0=0.067$ are displayed in
Figs.\,\ref{fig:snapshots_xz_amp067}--\ref{fig:snapshots_xz_yz_xy_amp067}.
Here, the nonaffine measure was computed for two consecutive
configurations at zero strain, and the time interval in
Eq.\,(\ref{Eq:D2min}) is set to $\Delta t = T$. For visualization of
irreversible rearrangements during one shear cycle, only atoms with
relatively large nonaffine displacements are shown, \textit{i.e.},
$D^2(n\,T, T)>0.04\,\sigma^2$, where $n$ is the integer. The typical
cage size at the density $\rho=1.2\,\sigma^{-3}$ is
$r_c\approx0.1\,\sigma$~\cite{Priezjev13}.  As shown in
Fig.\,\ref{fig:snapshots_xz_amp067}\,(a,\,b), the structural
relaxation during mechanical annealing at $\gamma_0=0.067$ (the blue
curve, $t \lesssim 1400\,T$ in Fig.\,\ref{fig:poten_amp067_3prs})
proceeds via rearrangement of relatively small clusters of atoms.
Interestingly, the shear band in
Fig.\,\ref{fig:snapshots_xz_amp067}\,(c) is only partially formed
along the $yz$ plane as the yielding transition occurs gradually
during several hundred periods (\textit{i.e.}, $1600\,T \lesssim t
\lesssim 2200\,T$ in Fig.\,\ref{fig:poten_amp067_3prs}). Finally,
the snapshot at Fig.\,\ref{fig:snapshots_xz_amp067}\,(d) confirms
that the shear band is fully developed after yielding, whereas
irreversible displacements outside the shear band are nearly absent
since the effective strain amplitude in that region is smaller than
$\gamma_0=0.067$.

\vskip 0.05in


The structural relaxation before and after the yielding transition
at $\gamma_0=0.067$ is similar in the case of alternating shear
along the $xz$ and $yz$ panes, as evident from
Fig.\,\ref{fig:snapshots_xz_yz_amp067}. Notice, however, that the
shear band is formed at $t=1600\,T$, since the sharp yielding
transition occurs already at $t\approx1400\,T$ (see
Fig.\,\ref{fig:poten_amp067_3prs}).  While the orientation of the
shear band in Figs.\,\ref{fig:snapshots_xz_amp067} and
\ref{fig:snapshots_xz_yz_xy_amp067} is difficult to predict, the
deformation protocol with alternating shear along the $xz$ and $yz$
planes imposes formation of the shear band along the perpendicular
plane (\textit{i.e.}, along the $xy$ plane).  Further, the
appearance of a nearly percolating cluster of mobile atoms in
Fig.\,\ref{fig:snapshots_xz_yz_xy_amp067}\,(a) precedes the yielding
transition when shear orientation is alternated in all three
dimensions.   Moreover, once the shear band is formed along the $xy$
plane, as illustrated in
Fig.\,\ref{fig:snapshots_xz_yz_xy_amp067}\,(b--d), every third cycle
with $\gamma_0=0.067$ along the $xy$ plane induces relatively large
irreversible rearrangements outside the shear band. It should also
be commented that only one shear band is formed at the largest
strain amplitude $\gamma_0=0.070$ for alternating shear along three
dimensions (not shown).   Taken together, the evolution of spatial
distributions of atoms with large nonaffine displacements during
cyclic loading at the critical strain amplitude, shown in
Figs.\,\ref{fig:snapshots_xz_amp067}--\ref{fig:snapshots_xz_yz_xy_amp067},
correlates well with the onset of yielding reported in
Fig.\,\ref{fig:poten_amp067_3prs}.

\section{Conclusions}

In summary, molecular dynamics simulations were carried out to
examine the influence of alternating shear orientation during
periodic shear of amorphous materials near the critical strain
amplitude.  We considered a binary glass rapidly quenched well below
the glass transition temperature and then periodically deformed
along a single plane or alternating planes in two or three spatial
dimensions.   It was found that at strain amplitudes below the
critical value, the glass continues exploring lower potential energy
states via collective, irreversible rearrangements of atoms.  By
contrast, at the critical strain amplitude and above, the structural
relaxation is followed by the yielding transition and formation of a
shear band along one of the planes. It was shown that the number of
transient cycles before yielding is reduced when an additional
alternation of shear orientation is included in the deformation
protocol at a given strain amplitude.

\vskip 0.05in

Interestingly, following the yielding transition, the glass outside
the shear band continues to relax, since the effective strain
amplitude in the solid domain is reduced when shear is applied along
a single plane or alternated between two planes.  In the case of
alternating shear orientation in all three dimensions, however,
cyclic shear during one of the three periods is applied along the
shear band, and the rest of the material is deformed with the
imposed strain amplitude, leading to a nearly constant level of the
potential energy.  Furthermore, the deformation protocol with
alternating shear along two planes imposes the formation of the
shear band along the perpendicular plane, while for other protocols
the orientation of the shear band cannot be easily predicted.  These
results are important for the development of processing methods that
require precise control of relaxation and yielding in amorphous
solids.

\section*{Acknowledgments}

Financial support from the National Science Foundation (CNS-1531923)
is gratefully acknowledged.  The article was prepared within the
framework of the HSE University Basic Research Program and funded in
part by the Russian Academic Excellence Project `5-100'. The
numerical simulations were carried out at Wright State University's
Computing Facility and the Ohio Supercomputer Center. The molecular
dynamics simulations were performed using the LAMMPS code developed
at Sandia National Laboratories~\cite{Lammps}.


%

%

%

%

%

%

%

%

\bibliographystyle{prsty}

\end{document}